\documentclass[aps,prb,twocolumn,showpacs,superscriptaddress]{revtex4-1}
\usepackage[mathscr]{eucal}
\usepackage{graphicx}
\usepackage{amsmath}
\usepackage{amstext}
\usepackage{amssymb}
\usepackage{amsfonts}
\usepackage{amsbsy}
\usepackage{dcolumn}
\usepackage{bm}
\usepackage{multirow}
\usepackage{color}
\usepackage{datetime}
\usepackage{hyperref}

\begin{document}

\title{Sign structure and ground-state properties for spin-$S$ $t$-$J$ chain}

\author{Qing-Rui Wang}
\affiliation{Institute for Advanced Study, Tsinghua University, Beijing 100084, China}
\author{Peng Ye}\email[]{pye@perimeterinstitute.ca}
\affiliation{Perimeter Institute for Theoretical Physics, Waterloo, Ontario, Canada, N2L 2Y5}
\date{{\currenttime, \small\today}}

\begin{abstract}
The antiferromagnetic Heisenberg spin chain of odd spin $S$ is in the Haldane phase with several defining physical properties, such as thermodynamical ground-state degeneracy, symmetry-protected edge states, and nonzero string order parameter. If nonzero hole concentration $\delta$ and hole hopping energy $t$ are considered, the spin chain is replaced by a spin-$S$ $t$-$J$ chain. The motivation of this paper is to generalize the discussions of the Haldane phase to the doped spin chain.
The \emph{first result} of this paper is that, for the model considered here, the $\mathbb{Z}_2$ sign structure in the usual Ising basis can be totally removed by two consecutive unitary transformations consisting of a spatially local one and a nonlocal one. Direct from the sign structure, the \emph{second result} of this paper is that the Marshall theorem and the Lieb-Mattis theorem for pure spin systems are generalized to the $t$-$J$ chain for arbitrary $S$ and $\delta$. A corollary of the theorem provides us with the ground-state degeneracy in the thermodynamic limit. The \emph{third result} of this paper is about the phase diagram. We show that the defining properties of the Haldane phase survive in the small $t/J$ limit. The large $t/J$ phase supports a gapped spin sector with similar properties (ground-state degeneracy, edge state, and string order parameter) of the Haldane chain, although the charge sector is gapless.
\end{abstract}

\maketitle


\section{Introduction}
It is well known, from Haldane's conjecture, that the one-dimensional (1D) spin-$S$ antiferromagnetic Heisenberg model is gapped if $S$ is integer, and gapless if $S$ is half-odd-integer \cite{Haldane1983,Haldane1985}. The odd integral spin chain is in the Haldane phase, which is one of the simplest examples of the symmetry-protected topological (SPT) phases \cite{chen2012,chen2013}.
The Haldane phase is characterized by several nontrivial defining properties: the ground-state degeneracy in open boundary condition, the symmetry-protected edge states, the hidden antiferromagnetic order unveiled by nonzero string order parameter, etc. The motivation of this paper is to generalize these discussions to the doped spin chain and investigate the physical properties of the ground state of the 1D $t$-$J$ model.

The Hamiltonian of the 1D $t$-$J$ model investigated in this paper is given by
\begin{align}\label{t-J}
 H= &\sum_{\langle ij\rangle,m} -t_{ij} (c_{i,m}^\dagger c_{j,m}+\mathrm{H.c.})\nonumber\\
 &+ \sum_{\langle ij\rangle} J_{ij} ( \mathbf S_i\cdot \mathbf S_j - S^2 n_i n_j ).
\end{align}
The system is on an open chain with $L$ sites and $N$ spin-$S$ ``electrons'' together with $N_h=L-N$ holes. The hole doping concentration is defined as $\delta\equiv N_h/L$. The ``electron'' annihilation operator $c_{i,m}$ at site $i$ carries $z$-component spin $S^z=m$ ($=-S,-S+1,..., S$), and can be either {\emph{hard-core} bosonic or fermionic. $n_i = \sum_m n_{i,m} = \sum_m c_{i,m}^\dagger c_{i,m}$  is the particle number operator of ``electrons'' at site $i$. The total Hilbert space $\mathcal{H}$ is defined by the non-doubly constraint ``$n_i\leq1$'' at \emph{every} site $i$.  $t_{ij}$ and $J_{ij}$ on each link $\langle ij \rangle$ are always positive, and the link dependence is allowed. In this paper, the first term and the second term in Eq.~(\ref{t-J}) will be called ``$t$-term'' $H_t$ and ``$J$-term'' $H_J$, respectively.
When the doping concentration $\delta$ is zero, the above $t$-$J$ model is reduced to the antiferromagnetic Heisenberg model.

From the perspective of numerical analysis, quantum Monte Carlo (QMC) is a complementary tool to investigate strongly correlated systems \cite{MC_book}. However, the notorious \emph{sign problem} of electronic and frustrated bosonic models generically plagues the application of QMC, for the functional integrals usually do not have a positive-definite measure \cite{MC1}. Although it is generally a nondeterministic polynomial hard problem \cite{MC_NP}, there exist some special models which turn out to be \emph{sign problem free}, after carefully analyzing the \emph{sign structure} of them \cite{MC2,MC3,MC4}. Here, sign structure is roughly referred to as in what circumstances the minus ``probabilities'' arise in the functional integral. Recently, Berg \textit{et al.} have also attempted to modify a fermionic model to make it sign problem free before using the QMC method \cite{Berg2012}.


The notion of \emph{sign structure} of a given model is also especially emphasized throughout this paper. 
The validity of many theorems, which have been playing very important roles in an analytic approach to universal properties of strongly correlated quantum magnets in low dimensions, crucially depends on the sign structure of these models.
Along with the celebrated Marshall theorem\cite{Marshall1955}\footnote{The Marshall theorem is not restricted only in systems with equal size sublattices and can be stated as follows: the ground state of the spin-$S$ antiferromagnetic Heisenberg model with $N_A(N_B)$ sublattice $A(B)$ sites has total spin quantum number $S\ |N_A-N_B|$. This conclusion can be drawn from the observation that the ground state of the infinite-range antiferromagnetic Heisenberg model (\emph{i.e.} $H_{\infty} = J \sum_{i\in A,j\in B} \bold S_i \cdot \bold S_j$ with $J>0$), constructed in the last step of the proof of the Marshall theorem, has total spin quantum number $S\ |N_A-N_B|$ \cite{Lieb1989,Auerbach}. This version of Marshall theorem is used in our paper to prove our generalized theorems.}, many other theorems on pure spin models have been proposed \cite{Lieb-Schultz-Mattis1961,Lieb-Mattis1962,Auerbach}.
However, if charge degrees of freedom are introduced by doping holes, problems generically become more intricate. Most of the rigorous results on doped spin models, \emph{e.g.}, Nagaoka theorem \cite{Nagaoka1966,Tasaki1989}, Lieb theorem \cite{Lieb1989}, flat-band ferromagnetism \cite{Mielke1,Mielke2,Mielke3,Tasaki1992,MielkeTasaki}, and Tasaki theorem \cite{Tasaki1990}, are valid merely for one hole doping or other fixed doping. We will generalize the Marshall theorem and the Lieb-Mattis theorem to the 1D $t$-$J$ model for arbitrary spin $S$ and hole density $\delta$ in this paper.

The organization of the paper is as follows. We first set up a general theory concerning the sign structure of a generic Hamiltonian and its ground state in Sec.~\ref{sec_sign_general}.
Following this theory, the sign structure of the 1D spin-$S$ $t$-$J$ model in the usual Ising basis is identified in Sec.~\ref{sec_sign_tJ}, by consecutively performing two unitary transformations: the ``Marshall sign transformation'' \cite{Marshall1955,Auerbach}, which is spatially local; and the ``phase string transformation'' originally proposed by Weng \emph{et al.} \cite{phase_1D_1,phase_1D_2,Weng1997,Wu2008sign}, which is spatially nonlocal \footnote{Nonlocal unitary transformation means that it cannot be decomposed to a product of several unitary transformations acting on only nearby sites.}.

In Sec.~\ref{sec_doping_Haldane}, we turn to the physical properties (rather than ``abstract'' sign structure) of the state after doping the Haldane phase. Two theorems (generalized Marshall theorem and generalized Lieb-Mattis theorem) are proposed in Sec.~\ref{subsec_GSD}, directly from the sign structure of the 1D spin-$S$ $t$-$J$ model. One corollary of the theorem is about the ground-state degeneracy in the thermodynamic limit.

In addition to rigorous analysis above, Sec.~\ref{sec_phase_diag} deals with physical properties of the $t$-$J$ model in other approaches. We first briefly discuss the phase diagram of the model.
Then, topological properties, such as edge states and string order parameter, are investigated. We also point out the relation between sign structure and string order parameter.

We summarize this paper in Sec.~\ref{sec_summary}. Some future directions based on this work will also be discussed.
One appealing direction is to investigate the interplay of spin rotation symmetry and topological properties (\emph{e.g.}, ground-state degeneracy and edge states) in the presence of holes by introducing spin-orbital couplings, while those holes play the role of a ``symmetry-breaker''.


\section{Sign structure of a generic model}
\label{sec_sign_general}

In this section, we emphasize the sign structure of a given Hamiltonian and the ground state. We show that it is the Perron-Frobenius theorem that connects the sign-problem-free condition in QMC and the condition for trivial sign structure of the ground state.
The reason for identifying the sign structure of the model is that, many key physical properties of the model, such as the ground-state total spin and ground-state degeneracy, are determined by the sign structure.
The trivial sign basis is also constructed explicitly and a geometric interpretation for the sign trivial condition is given.

\subsection{Sign structure of a Hamiltonian in QMC}
We first clarify in what circumstances we can say a model is \emph{sign problem free} in QMC. The starting point of QMC is to express the partition function as ($\beta={1}/{T}$ with temperature $T$)
\begin{equation}\label{MC2}
  \mathsf{Z} \equiv \mathrm{Tr}\, e^{- \beta H} = \sum_{k=0}^\infty \frac{\beta^k}{k!} \sum_{\{\alpha_i\}} \prod_{i=0}^{k-1} \langle \alpha_{i+1} | (-H) | \alpha_i \rangle,
\end{equation}
where $| \alpha_i \rangle \in \mathsf{\Gamma}$ and $\mathsf{\Gamma}$ is a basis we choose. We assume all the matrix elements of the Hamiltonian in the basis $\mathsf{\Gamma}$ are real numbers. This expression is called stochastic series expansion \cite{SSE1,SSE2}. The partition function $\mathsf{Z}$ can be expressed as a summation of \emph{non-negative} numbers, labeled by integer $k$ and a sequence of basis states $\{|\alpha_i\rangle\}$ with $|\alpha_0\rangle=|\alpha_k\rangle$, if
\begin{equation}\label{sign_free2}
  \prod_{i=0}^{k-1} \langle \alpha_{i+1} | (-H) | \alpha_i \rangle \geq 0,\quad \forall\ k\geq 2,\ |\alpha_i\rangle \in \mathsf{\Gamma}, \ |\alpha_0\rangle=|\alpha_k\rangle.
\end{equation}
The $k=1$ term $\langle\alpha_0|(-H)|\alpha_0\rangle$ is omitted, because it can be shifted to be positive by adding a constant term to the Hamiltonian. A \emph{geometric interpretation} of Eq.~(\ref{sign_free2}) based on the notion of \emph{state complex} will be given in Sec.~\ref{subsec_geo}.

Note that we can also use the world-line QMC expression
\begin{eqnarray}\label{MC1}\nonumber
  \mathsf{Z} &=& \mathrm{Tr}\, e^{- \beta H}
    \simeq \sum_{ \{\alpha_i\} } \prod_i \langle \alpha_{i+1} | e^{-\Delta \tau H} | \alpha_i \rangle \\
    &\simeq& \sum_{ \{\alpha_i\} } \prod_i \left[ \delta_{\alpha_{i+1},\alpha_i} + \Delta \tau \langle \alpha_{i+1} | (-H) | \alpha_i \rangle \right],
\end{eqnarray}
to derive the condition (\ref{sign_free2}): to employ the standard MC method, one should make sure that the ``probability" in Eq.~({\ref{MC1}}) is non-negative, which implies Eq.~(\ref{sign_free2}).

\subsection{Sign structure of the ground state}
\label{subsec_GSsign}
In this paper, when talking about the \emph{sign structure} of a state
\begin{equation}\label{}
  |\Psi\rangle = \sum_{\alpha \in \mathsf{\Gamma}} a_\alpha |\alpha\rangle
\end{equation}
in a given basis $\mathsf{\Gamma}$, we are referring to the signs of the coefficients $a_\alpha$ in this basis (we assume the Hamiltonian matrix elements are all real, then all energy eigenstates have real coefficients in this basis). The sign structure of the ground state of a model is closely related to the sign structure of the Hamiltonian. The bridge connecting these two sign structures is the Perron-Frobenius theorem.

We first present the Perron-Frobenius theorem \cite{Berman}: let $A=\{a_{ij}\}$ be an $n\times n$ matrix with $a_{ij}\leq0$ for $i\neq j$. If $A$ is irreducible in the sense that, for any $i\neq j$, there exists a positive integer $k$, such that $(A^k)_{ij}\neq 0$, then the Perron-Frobenius theorem states that the eigenvector of $A$ with minimum eigenvalue is unique, and has strictly positive coefficients in this basis.

Physically, for a given Hamiltonian $H$ and given basis $\mathsf{\Gamma}_0$, the conditions for the Perron-Frobenius theorem are (i) the inequality
\begin{equation}\label{sign_free1}
  \langle\alpha | H |\beta\rangle \leq 0, \quad \forall\ |\alpha\rangle, |\beta\rangle \in \mathsf{\Gamma}_0,\ |\alpha\rangle \neq |\beta\rangle.
\end{equation}
is satisfied; (ii) the model is irreducible, in the sense that every two states $|\alpha\rangle$ and $|\beta\rangle$ in $\mathsf{\Gamma}_0$ can be connected by a consecutive action of the Hamiltonian. In more precise words, there exists a sequence of basis states $\{|\alpha_1\rangle, |\alpha_2\rangle, ..., |\alpha_K\rangle\}$ with $|\alpha_1\rangle = |\alpha\rangle$ and $|\beta\rangle = |\alpha_K\rangle$, such that $\langle\alpha_i|H|\alpha_{i+1}\rangle \neq 0$ for all $0 \leq i \leq K-1$. The Perron-Frobenius theorem states that the ground state
\begin{equation}\label{}
  |\Psi_0\rangle = \sum_{\alpha \in \mathsf{\Gamma}} a_\alpha |\alpha\rangle
\end{equation}
is unique, and the coefficients satisfy
\begin{equation}\label{sign_wf}
  a_\alpha > 0,\quad \forall |\alpha\rangle \in \mathsf{\Gamma}_0.
\end{equation}

The Perron-Frobenius theorem makes it possible to unveil the sign structure of the ground state wave function (\ref{sign_wf}) from the sign structure of the Hamiltonian (\ref{sign_free1}). Since the coefficients $a_\alpha$ are all positive, we will say a model is accompanied with \emph{trivial sign structure} in the basis $\mathsf{\Gamma}_0$ if Eq.~(\ref{sign_free1}) is satisfied.

As a matter of fact, the two conditions on the sign structure of the Hamiltonian equations (\ref{sign_free2}) and (\ref{sign_free1}) are equivalent to each other: if Eq.~(\ref{sign_free2}) is satisfied in the basis $\mathsf{\Gamma}$, then there exists a basis $\mathsf{\Gamma}_0$, which is obtained from $\mathsf{\Gamma}$ by a unitary transformation, such that Eq.~(\ref{sign_free1}) is satisfied (the explicit construction of $\mathsf{\Gamma}_0$ from $\mathsf{\Gamma}$ is given in Appendix~\ref{Appen_sign}); the converse proposition is obviously true. Since the unitary transformation from $\mathsf{\Gamma}$ to $\mathsf{\Gamma}_0$ can remove the minus signs of the ground-state wave function in the basis $\mathsf{\Gamma}$, we can also say the model has \emph{removable $\mathbb{Z}_2$ sign structure} in the basis $\mathsf{\Gamma}$ if Eq.~(\ref{sign_free2}) is satisfied.

\subsection{Geometric interpretation}
\label{subsec_geo}

The condition for \emph{removable $\mathbb{Z}_2$ sign structure} Eq.~(\ref{sign_free2}) can be interpreted geometrically.

A graph can be used to represent the structure of a given model (see Fig.~\ref{fig:Graph}): each basis state in $\mathsf{\Gamma}$ is denoted by a point; each line connecting two points $|\alpha\rangle$ and $|\beta\rangle$ represents the nonzero matrix element $\langle\alpha|(-H)|\beta\rangle$. We will call this graph a \emph{state complex}, since it is a simplicial complex in which the 0-simplices represent the basis states of the total Hilbert space $\mathcal{H}$, rather than points in space or spacetime in the group cohomology classification theory of bosonic SPT states \cite{GW09,Pollmann10,chen2012,chen2013}.

\begin{figure}[t]
\includegraphics[width=8.5cm]{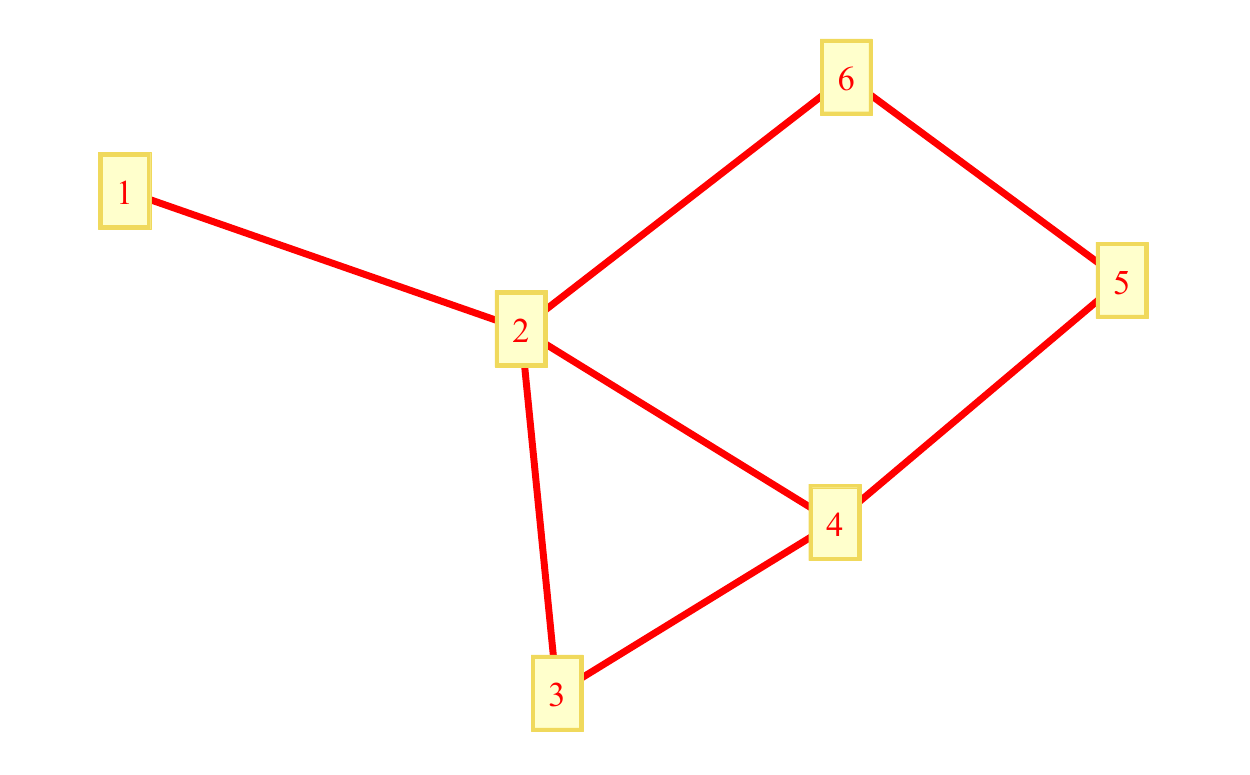}
\caption{\label{fig:Graph}(Color online). State complex. An example of state complex with six basis states is shown in this figure. Each point $i$ represents a basis state $|\alpha_i\rangle$; each line connecting two points $i$ and $j$ represents nonzero number $\langle \alpha_i | (-H) | \alpha_j \rangle$. The geometric interpretation of Eq.~(\ref{sign_free2}) is that the flux through each closed loop in the state complex is non-negative. }
\end{figure}

By viewing $\langle\alpha|(-H)|\beta\rangle$ as Berry connection, Eq.~(\ref{sign_free2}) means the flux through any closed loop of the graph is positive. Mathematically, it suggests that the Hamiltonian, in the basis $\mathsf{\Gamma}$ as a 1-cochain, belongs to the trivial element of the first cohomology group of the state complex. At the same time, the condition for trivial sign structure Eq.~(\ref{sign_free1}) means the connections are all positive, which also indicates that all the fluxes are positive. It is for this reason that the two conditions (\ref{sign_free2}) and (\ref{sign_free1}) are equivalent.


The notion of state complex can be used to construct a family of exactly solvable models. In general, we can construct a fixed-point wave function with trivial $\mathbb{Z}_2$ sign structure: all amplitudes are the same in an irreducible space. The Hamiltonian has the following properties: all off-diagonal elements are $-1$ or $0$; the sum of every column of the Hamiltonian matrix is zero.
We call it the \emph{Laplacian model} following the spectral graph theory, since the Hamiltonian is the Laplacian operator of a graph with every vertex representing a basis state of the Hilbert space. The ground state of the model is an equal weight superposition of all basis states in this irreducible space. It is a fixed-point wave function in the sense that the amplitudes are either one or zero.

Many exactly solvable models, of which the ground state is an equal weight superposition of a
collection of basis states, belong to the category of the Laplacian model.
For example, the Rokhsar-Kivelson (RK) point of the quantum dimer model \cite{RK1988}, the toric code model \cite{Kitaev2003}, and the doubled semion model \cite{string_net2005,doubled_semion2012} all possess ground states with the above structure.

\section{Sign structure of the 1D spin-$S$ $t$-$J$ model}
\label{sec_sign_tJ}
After clarifying the sign structure of a generic model and its ground state, we now turn to the sign structure of the 1D $t$-$J$ model for arbitrary spin $S$. As can be verified, Eq.~(\ref{sign_free2}) is satisfied for the 1D spin-$S$ $t$-$J$ model in the usual Ising basis. Therefore, our next task is to find out the special basis $\mathsf{\Gamma}_0$ by properly adding minus signs to the usual Ising basis states, in order to support the trivial sign structure of the model.

In this section, we will first unveil the sign structure of the 1D $t$-$J$ model from the basis transformation point of view. Although expressions in fractionization formalism are sometimes tedious, the transformation of the Hamiltonian, discussed in Appendix~\ref{Appen_transf_of_H}, is an equivalent approach to show the trivial sign structure of the model directly. After that, a lemma on the sign structure of the ground state of the model is proposed. The sign structure of the (doped) Affleck-Kennedy-Lieb-Tasaki (AKLT) state is analyzed in the last subsection.

\subsection{Basis transformation}
\label{subsec_basis_transf}
The trivial sign basis $\mathsf{\Gamma}_0$ will be constructed from the usual Ising basis by two consecutive unitary transformations: spatially local Marshall sign transformation (MST) \cite{Marshall1955,Auerbach} and spatially nonlocal phase string transformation (PST) \cite{phase_1D_1,phase_1D_2,Weng1997,Wu2008sign}.

\subsubsection{Marshall sign transformation}
The usual Ising basis of the Hilbert space of the 1D $t$-$J$ model can be denoted by
\begin{equation}\label{}
  |\{h_i\};\{m_i\}\rangle = c_{i_1,m_{i_1}}^\dagger c_{i_2,m_{i_2}}^\dagger ... \ c_{i_N,m_{i_N}}^\dagger |0\rangle,
\end{equation}
where $\{h_i\}$ and $\{m_i\}$ represent the positions of the holes and the $z$-component spin of the ``electrons'', respectively. We choose the order of the creation operators such that $i_1<i_2<...<i_N$. The set $\{1,2,...,L\}$ is the disjoint union of two sets $\{i_1,i_2,...,i_N\}$ and $\{h_1,h_2,...,h_{N_h}\}$.

As a spatially \emph{local} unitary transformation, MST transforms the usual Ising basis to the \emph{Marshall basis} \cite{Marshall1955,Auerbach}:
\begin{equation}\label{MarshallBasis}
  |\{h_i\};\{m_i\}\rangle' = \prod_{j,m} (-1)^{j (S+m) n_{j,m}} |\{h_i\};\{m_i\}\rangle\,.
\end{equation}
In this new basis, the usual Ising basis state $|m_j\rangle$ on sites $j$ is modified by an additional sign $(-1)^{S+m_j}$ if $j$ belongs to sublattice $B$, and unchanged if $j$ belongs to sublattice $A$.

The $J$-term in the new basis has only non-positive off-diagonal matrix elements. However, there is a price to be paid because a new sign $(-1)^{S+m}$ appears in front of the $t$-term in the Marshall basis. In order to further remove the new signs in front of the $t$-term, we should continue to perform the so-called phase string transformation.

\subsubsection{Phase string transformation}
The purpose of PST \cite{phase_1D_1,phase_1D_2,Weng1997,Wu2008sign} is to absorb the signs in the $t$-term while keeping the $J$-term invariant. We can introduce a sign $(-1)^{S+m}$ whenever a hole appears on the left side of a spin with $S^z=m$. This nonlocal unitary transformation results in a basis rotation from the Marshall basis Eq.~(\ref{MarshallBasis}) to the \emph{rotated Ising basis}:
\begin{equation}\label{rotatedIsing}
  |\{h_i\};\{m_i\}\rangle'' = \prod_{i<j;m} (-1)^{(S+m) n^h_{i} n_{j,m}} |\{h_i\};\{m_i\}\rangle',
\end{equation}
where $n^h_i = 1 - n_i$ is the hole number operator.

As claimed above, PST successfully makes the $t$-term sign trivial  without introducing new signs in front of the $J$-term. Therefore, MST and PST together (\emph{i.e.} a total unitary transformation $U=U_{\mathrm{pst}}\ U_{\mathrm{mst}}$) make the Hamiltonian Eq.~(\ref{t-J}) sign trivial in a new basis.

\subsection{Sign structure of the ground state}
\label{subsec_tJ_GS}
After clarifying the sign structure of the Hamiltonian of the 1D spin-$S$ $t$-$J$ model by two unitary transformations, the sign structure of the ground state can be identified following the discussions in Sec.~\ref{subsec_GSsign}. We can now draw one of the key conclusions in this paper on the 1D spin-$S$ $t$-$J$ model Eq.~(\ref{t-J}):\\

\emph{Lemma. In any subspace labeled by $z$-component total spin quantum number $S^z_\mathrm{tot}$, the lowest energy state is unique and has positive definite coefficients in the rotated Ising basis [see Eq.~(\ref{rotatedIsing})].}\\

The proof is as follows. We have shown the following in Sec.\ref{subsec_basis_transf}: (i) in the rotated Ising basis Eq.~(\ref{rotatedIsing}), the off-diagonal matrix elements of the 1D $t$-$J$ model are all non-positive: $\langle \alpha | H |\beta\rangle \leq 0$ for all states $|\alpha\rangle \neq |\beta\rangle$. Now, let us fix the $z$ component of the total spin $S^z_{\mathrm{tot}}=M$. One can also verify that (ii) the space $\mathcal{H} ( S^z_{\mathrm{tot}}=M )$, as a subspace of the total Hilbert space $\mathcal{H}$, is closed and irreducible under the action of the Hamiltonian Eq.~(\ref{t-J}). Provided with the two properties of our model, we can state now, due to the {Perron-Frobenius theorem}, that the ground state in the subspace $\mathcal{H} ( S^z_{\mathrm{tot}}=M )$ is unique and has strictly positive coefficients in the rotated Ising basis.

\subsection{Sign structure of (doped) AKLT state}
\label{subsec_AKLT_sign}
Although the trivial sign structure of a Hamiltonian leads to the trivial sign structure of the ground state, the converse proposition is not true.

The AKLT model \cite{AKLT1987,AKLT1988} is a simple counterexample. The Hamiltonian of the AKLT model
\begin{eqnarray}\label{AKLT_H}
    H_{\mathrm{AKLT}} = J\sum_{\langle ij \rangle} \left(\mathbf S_i\cdot \mathbf S_{j} + \frac{1}{3} \left( \mathbf S_i\cdot \mathbf S_{j} \right)^2 + \frac{2}{3} \right)
\end{eqnarray}
has a biquadratic term, which makes the model not sign trivial in the usual Ising basis. The nontrivial sign structure can be verified by noticing
\begin{equation}\label{AKLT_sign}
  \langle 00|(-H_{ij})|-+\rangle \langle -+|(-H_{ij})|+-\rangle \langle +-|(-H_{ij})|00\rangle < 0,
\end{equation}
where $|+\rangle,|0\rangle,|-\rangle$ are three states with $S^z=1,0,-1$, respectively, and $H_{ij}$ is the term in Eq.~(\ref{AKLT_H}) involving the degrees of freedom on the link $\langle ij \rangle$. According to the criterion for removable $\mathbb{Z}_2$ sign structure Eq.~(\ref{sign_free2}), the negative sign in Eq.~(\ref{AKLT_sign}) indicates there does not exist a sign attachment procedure to the usual Ising basis, such that the AKLT Hamiltonian becomes sign trivial, \emph{i.e.}, Eq.~(\ref{sign_free1}) is satisfied. The condition for the Perron-Frobenius theorem is broken, but the conclusion of the theorem will be shown to be also true in the Marshall basis.

The ground state of the AKLT model
\begin{eqnarray}\label{AKLT_wf}
    |\Psi_{\mathrm{AKLT}}\rangle = \prod_i \left( b_{i\uparrow}^\dagger b_{i+1\downarrow}^\dagger - b_{i\downarrow}^\dagger b_{i+1\uparrow}^\dagger \right) |0\rangle
\end{eqnarray}
has trivial sign structure in the Marshall basis: by performing the Marshall sign transformation $b_{j\sigma} \rightarrow (-\sigma)^j b_{j\sigma}$, the AKLT state becomes a state with only non-negative amplitude in the Marshall basis Eq.~\ref{MarshallBasis}.
It means the sign structure of the AKLT state is exactly the same as that of the spin-$1$ antiferromagnetic Heisenberg model, even though the sign structures of the two Hamiltonians are different.

The doped AKLT state \cite{Zhang1989} has similar sign structure. In Sec.~\ref{subsec_string_order}, we will discuss more about the relation between sign structure and string order parameter, and conclude that a state with nonzero string order parameter necessarily possesses this kind of sign structure.

\section{Doping Haldane phase}
\label{sec_doping_Haldane}
It is known from Haldane's conjecture that the antiferromagnetic Heisenberg chain with integral spin is gapped, while the half-odd-integral spin chain is gapless \cite{Haldane1983,Haldane1985}. The antiferromagnetic Heisenberg chain with odd spin is in the Haldane phase, which is one of the simplest examples of SPT phases in 1D \cite{chen2012,chen2013}.

In this section, we will dope the antiferromagnetic spin chain and investigate the ground-state properties of the $t$-$J$ model. Directly from the sign structure of the model emphasized above, exact theorems on ground-state total spin and degeneracy are discussed first. Then, we turn to other results of the $t$-$J$ model, including phase diagram and other topological properties (edge states and string order parameter).

\subsection{Ground state degeneracy}
\label{subsec_GSD}
The sign structure of the ground state of the 1D spin-$S$ $t$-$J$ model is given in Sec.~\ref{subsec_tJ_GS}. The sign structure is closely related to the physical properties of the model. In fact, we can prove the following two theorems (generalized Marshall theorem and generalized Lieb-Mattis theorem) for the 1D spin-$S$ $t$-$J$ model Eq.~(\ref{t-J}):\\

\emph{Theorem-1 (generalized Marshall)\label{thm1}
The ground state has total spin quantum number $S_{\mathrm{tot}}^0=0$ if $N$ is even, and $S_{\mathrm{tot}}^0=S$ if $N$ is odd. The ground state is unique apart from the $(2S_{\mathrm{tot}}^0+1)$-fold spin degeneracy.
}\\

\emph{Theorem-2 (generalized Lieb-Mattis)\label{thm2}
If we denote the lowest energy eigenvalue belonging to total spin $S_{\mathrm{tot}}$ by $E(S_{\mathrm{tot}})$, then the energy levels are ordered as $E(S_{\mathrm{tot}}) < E(S_{\mathrm{tot}}+1)$ for all $S_{\mathrm{tot}} \geq S_{\mathrm{tot}}^0$.
}\\


The proofs of the above two theorems, which directly rely on the sign structure of the ground state discussed in Sec.~\ref{subsec_tJ_GS}, are given in Appendix~\ref{Appen_theorems}.

Theorem-1 asserts that, for a finite system, the ground state total spin quantum number is $0$ or $S$, \textit{i.e.} the ground-state degeneracy is $1$ or $2S+1$, depending on the parity of the ``electron'' number.
The parity will become irrelevant in the thermodynamic limit. Thus there is a direct corollary of Theorem-1:\\

\emph{Corollary.
The ground-state degeneracy for the 1D spin-$S$ $t$-$J$ model Eq.~(\ref{t-J}) in the thermodynamic limit is at least $2S+2$.
}\\

To illustrate this result, numerical evidences of the ground-state total spin oscillation for small lattice size systems are shown in Fig.~\ref{fig:GSD}.

\begin{figure}[t]
\includegraphics[width=8cm]{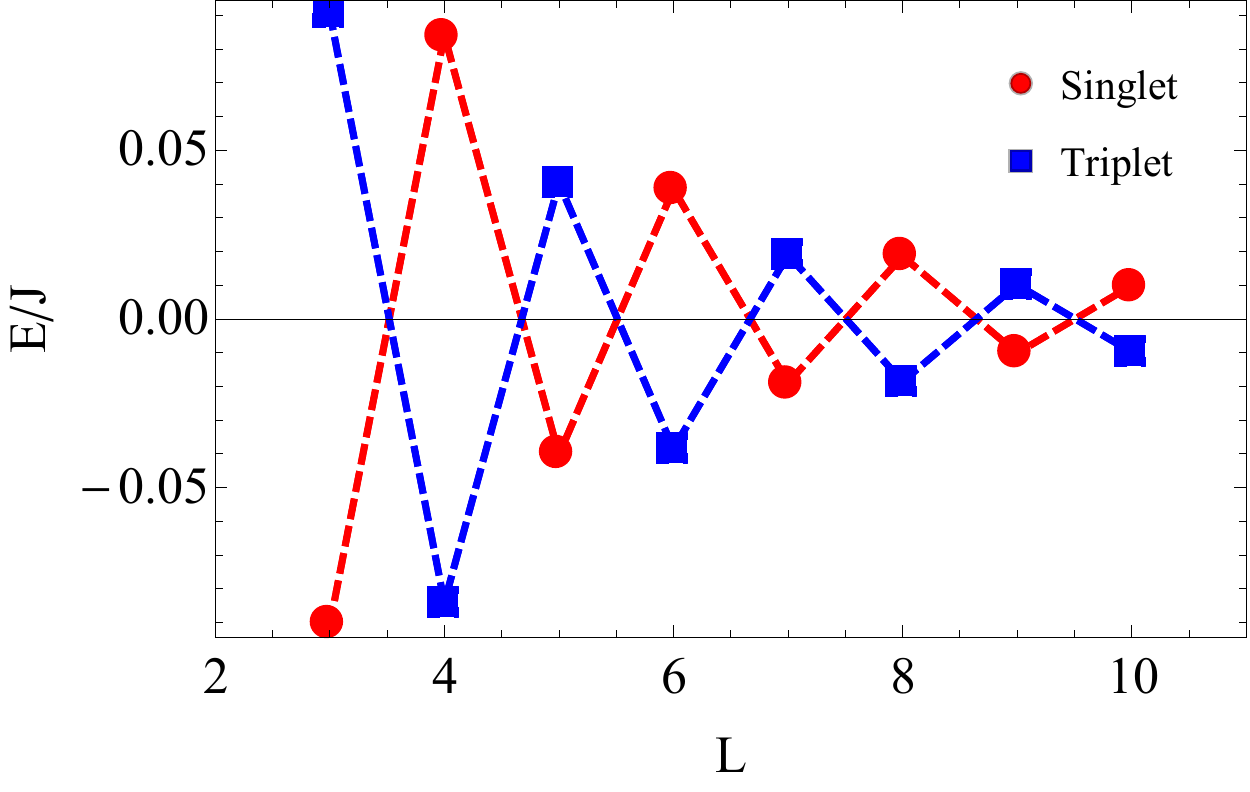}
\caption{\label{fig:GSD}(Color online). Ground state total spin oscillation. For the 1D spin-$1$ $t$-$J$ model ($t=10J$) with one hole ($N_h=1$), lowest energies (per site) of singlet states (red dots) and triplet states (blue squares) are plotted. At each $L$ the average of the two data points is set to zero for convenience. If $L$ is odd (even), \emph{i.e.} $N=L-N_h$ is even (odd), the singlet energy is lower (higher) than that of the triplet. These results agree with the generalized Marshall theorem and the corollary that the thermodynamical ground-state degeneracy is at least $2S+2$.}
\end{figure}

Note that the above theorems and corollary on the ground-state degeneracy are valid for any spin $S$, parameter region $t/J$ (recall that $t,J>0$), and hole doping concentration $\delta$. Different from the degeneracy from symmetry, this degeneracy is approximate for finite system and only exact in the thermodynamic limit. This property reminds us of the four fold ground-state degeneracy of the spin-$1$ Haldane chain. Our theorem indicates that, after doping holes to the Haldane chain, the ground state of the spin-$1$ $t$-$J$ chain may also possess some nontrivial topology.

We will investigate more about the topological properties (edge states and string order parameter) in the following section.

\subsection{Phase diagram, edge states and string order parameter}
\label{sec_phase_diag}
Similar to the phase diagram of the 1D spin-$1/2$ $t$-$J$ model \cite{phase_diag}, the spin-$S$ $t$-$J$ chain also has (at least) two phases. In Appendix~\ref{Appen_phase_diag}, we analyze the charge sector and spin sector of the model. The results are shown in Table~\ref{tab:phase_diag} and can be summarized as follows:

The charge sector of the 1D spin-$S$ $t$-$J$ model is gapped for small $t/J$, and gapless for large $t/J$, regardless of the integral or half-odd-integral nature of spin $S$. Meanwhile, the spin sector is gapped for integral $S$, and gapless for half-odd-integral $S$ in both small and large $t/J$ limit.

Now let us turn to the topological properties of edge states and string order parameter in different phases. For simplicity, we focus on spin-$1$ $t$-$J$ model henceforth.


\begin{table}[t]
\caption{\label{tab:phase_diag}
Small $t/J$ phase and large $t/J$ phase for spin integer and half-odd-integer $t$-$J$ chain.
}
\begin{center}
\begin{tabular}{|c|c||c|c|}
\hline
\multicolumn{2}{|c||}{} & \textbf{small} $t/J$ & \textbf{large} $t/J$ \\
\hline\hline
\multicolumn{1}{|c|}{\multirow{2}{*}{$S=1,2,3,...$} } & charge & gapped & gapless \\
\cline{2-4}
\multicolumn{1}{|c|}{\multirow{2}{*}{} } & spin & \multicolumn{2}{c|}{gapped} \\
\hline
\multicolumn{1}{|c|}{\multirow{2}{*}{$S=\frac{1}{2},\frac{3}{2},\frac{5}{2},...$} } & charge & gapped & gapless \\
\cline{2-4}
\multicolumn{1}{|c|}{\multirow{2}{*}{} } & spin & \multicolumn{2}{c|}{gapless} \\
\hline
\end{tabular}
\end{center}
\end{table}

\subsubsection{Edge states}
\label{subsec_edge_states}
One of the key properties characterizing the Haldane phase is the edge states of the spin-$1$ open chain. From the perspective of the AKLT model, the four fold ground-state degeneracy also comes from the edge states. The result in Sec.~\ref{subsec_GSD}, that the ground-state degeneracy is four even after doping the Haldane chain, suggests that there are also edge states for the $t$-$J$ chain.

In the small $t/J$ limit (see Appendix~\ref{Appen_phase_diag} for this phase), holes gather near the boundaries. If $N_L$ holes are at the left boundary of the chain, then there are $N_h-N_L$ holes at the right boundary.
Because of the inequality $0 \leq N_L \leq N_h$, apart from the spin degeneracy, there are $N_h+1$ ground states due to different hole distributions in the thermodynamic limit.
To connect any two of them by consecutive actions of the local hole hopping term, we need to perform at least $L$ times.
Therefore, the energy barrier between any two of these states is thermodynamically large, and they are degenerate in the thermodynamic limit. The key point is that the bulk of the chain is charge gapped for small $t/J$. Each of the $N_h+1$ ground states can be adiabatically connected to the Haldane chain with length $N$ through decreasing $t/J$ to 0.
Thus, the spin-$1/2$ edge states, supporting projective representation of SO(3) spin rotation, are also present for the spin-$1$ $t$-$J$ chain in the small $t/J$ limit.

On the other hand, in the large $t/J$ limit (see also Appendix~\ref{Appen_phase_diag} for this phase), mobile holes will smear the edge states. Because the ground state wave function has nonzero overlap with the basis state in which the hole distribution can be any possible one, the width of the edge state is of order $N_h$.
If $\delta<0.5$, the two edge states at two boundaries are still well separated in space. In this sense, the edge states are also stable against small perturbations as in the pure spin Haldane chain.

In summary, the small $t/J$ phase of the spin-$1$ $t$-$J$ chain can be adiabatically connected to the Haldane phase by decreasing $t/J$ to zero. Although the edge states of the Haldane chain may be smeared, they still exist after doping.
For the large $t/J$ phase, even though the charge sector is gapless, the edge states are also stable if $\delta<0.5$.

\subsubsection{String order parameter}
\label{subsec_string_order}
Aside from edge states, we can also use the string order parameter to unveil the nontrivial topology of the ground state. String order parameter, which is defined as
\begin{equation}\label{}
O_{\mathrm{string}} = \lim_{|i-j| \rightarrow \infty} \left\langle -S_i^z \exp\left( i \sum_{i<l<j} \pi S_l^z \right) S_j^z \right\rangle,
\end{equation}
was initially introduced to unveil the hidden order of the AKLT state and the Haldane phase \cite{deNijs1989,Girvin1989}. The small $t/J$ phase is the same as the Haldane phase. Therefore, we will mainly focus on the large $t/J$ phase and use string order parameter to detect the hidden order of it.

It seems impossible to analytically calculate the string order parameter for the large $t/J$ ground state of $t$-$J$ model which can not be written explicitly. But, we can rigorously calculate the string order parameter for the doped AKLT model \cite{Zhang1989}, which belongs to the same universality class as the ground state of the large $t/J$ limit of the spin-$1$ $t$-$J$ chain.

In fact, the string order transformation \cite{deNijs1989,Girvin1989} has two effects: keeping track of the sign structure and the antiferromagnetic order.
This can be seen from the action of string order transformation from Eq.~(\ref{AKLT_wf2}) to (\ref{AKLT2}) and the discussion around Eq.~(\ref{AKLT2}). In other words, the nonzero string order parameter in the Haldane phase indicates the rigidity of the Marshall sign structure and the antiferromagnetic order. A state with nonzero string order parameter necessarily possesses this kind of sign structure. Therefore, similar to the symmetry-breaking order, the sign structure of the ground state seems also to be a \emph{universal} property to characterize a generic phase.

In Appendix~\ref{Appen_stringorder}, we show that the string order parameter for spin-$1$ doped AKLT state is
\begin{equation}
  O_{\mathrm{string}}^{\mathrm{\emph{d}AKLT}} = \left(\frac{2}{3}\right)^2(1-\delta)^2.
\end{equation}
Note that when $\delta=0$, the above result becomes $(2/3)^2$, which is exactly the string order parameter for the AKLT state \cite{deNijs1989,Girvin1989}. This result suggests the system still possesses nontrivial topological structure as the Haldane chain as long as $\delta \neq 1$. The hidden antiferromagnetic order and the hidden $\mathbb{Z}_2 \times \mathbb{Z}_2$ symmetry-breaking theory can also be used to characterize this gapless phase.

\section{Summary}
\label{sec_summary}


To sum up, we have showed that, for any sign-problem-free model in QMC, one can find a special basis $\mathsf{\Gamma}_0$ in which the ground state has only non-negative coefficients. For the 1D spin-$S$ $t$-$J$ model, we can rigorously identify the sign structure after two unitary transformations: the Marshall sign transformation and the phase string transformation. After that, we proved two theorems concerning the ground state total spin quantum number and the ordering of excited states for this model, which determine the ground-state degeneracy in the thermodynamic limit. We also discussed the edge states and the string order parameter for the spin-$1$ $t$-$J$ chain. The small $t/J$ phase is gapped and can be adiabatically connected to the Haldane phase. The large $t/J$ phase has gapless charge excitation, but still possesses nontrivial topology as the Haldane chain, which is indicated by ground-state degeneracy, edge states and the nonzero string order parameter.

It is significant to look for the basis $\mathsf{\Gamma}_0$ for other lattice models with removable $\mathbb{Z}_2$ sign structure, and then analyze their properties following the same strategy in this paper.
One may also expect that in some models, generalizing the Abelian transformations (\emph{e.g.}, $U_{\rm mst}$ and $U_{\rm pst}$) to non-Abelian ones is desirable. Finally, doped holes in Eq.~(\ref{t-J}) do not break spin rotation symmetry. It is interesting to numerically and analytically investigate the relation between the topological properties (\emph{e.g.}, ground-state degeneracy, edge states) and spin rotation symmetry [SO(3) for integer spin; SU(2) for half-odd spin] through introducing spin-orbital couplings, \emph{i.e.}, replacing $t_{ij}c^\dagger_{i,m}c_{j,m}$ by a general term $t^{m,m'}_{ij}c^\dagger_{i,m}c_{j,m'}$ in Eq.~(\ref{t-J}).

\section*{Acknowledgement}
 We would like to thank Z.-Y. Weng, Y. Qi, and S.-S. Lee for many insightful discussions, and especially thank H. Yao for critical suggestions on our manuscript. We also thank Z.-X. Liu and X.-G. Wen for introducing the theory of SPT  from which our work was initiated. Q.R.W. is supported by NBRPC (Grants No. 2009CB929402 and No. 2010CB923003). P.Y. acknowledges the hospitality of the Institute for Advanced Study in Tsinghua University during 2013 Summer Forum  ``{the Interplay of Symmetry and Topology}''. Research at Perimeter Institute is supported by the Government of Canada through Industry Canada and by the Province of Ontario through the Ministry of Research and Innovation (P.Y.).

\appendix

\section{Construction of the trivial sign basis $\mathsf{\Gamma}_0$}
\label{Appen_sign}
In this appendix, we will \emph{remove} the $\mathbb{Z}_2$ signs of the basis $\mathsf{\Gamma}$ and construct the trivial sign basis $\mathsf{\Gamma}_0$ in Eq.~(\ref{sign_free1}) explicitly, provided that the condition for removable $\mathbb{Z}_2$ sign structure [Eq.~(\ref{sign_free2})] is satisfied in the basis $\mathsf{\Gamma}$.

Let us denote the dimension of the total Hilbert space $\mathcal{H}$ by $D$. The basis states in $\mathsf{\Gamma}$ will be denoted by $|\alpha_i\rangle$, \emph{i.e.}, $\mathsf{\Gamma} = \{ |\alpha_1\rangle, |\alpha_2\rangle, ... , |\alpha_D\rangle \}$. To simplify our notation, we define $A_{ij}=\langle \alpha_i | (-H) | \alpha_j \rangle$.

The construction of $\mathsf{\Gamma}_0$ contains several steps:

\emph{Step-1}.---Choose any two different states $|\alpha_{i_0}\rangle,|\alpha_{i_1}\rangle \in \mathsf{\Gamma}$, such that $A_{i_1 i_0}<0$. Note that if there do not exist such two states, then $\mathsf{\Gamma}_0=\mathsf{\Gamma}$ is the basis we want.

\emph{Step-2}.---Add a minus sign to $|\alpha_{i_1}\rangle$, \emph{i.e.} $|\alpha_{i_1}\rangle \longrightarrow -|\alpha_{i_1}\rangle$. We will denote the new basis state by $|\alpha_{i_1}\rangle$ thereafter. After this transformation, $A_{i_1 i_0}>0$ is satisfied. Now proceed to one of the following two steps.

\emph{Step-3}.---If there exists a basis state $|\alpha_{i_2}\rangle$, such that $A_{i_2 i_1}<0$, then back to step-2 and do similar transformation to $|\alpha_{i_2}\rangle$. The basis state $|\alpha_{i_2}\rangle$ must be different from all the previously visited states, otherwise there exists a sequence of states from $|\alpha_{i_2}\rangle$ to $|\alpha_{i_2}\rangle$ itself, such that Eq.~(\ref{sign_free2}) is broken.

\emph{Step-4}.---If for any basis state $|\alpha_{i_2}\rangle$, the condition $A_{i_2 i_1} \geq 0$ is always satisfied, then the present loop is finished and we should back to step-1 to start a new loop.

If a loop is finished, we get an integer sequence $i_0,i_1,...,i_n$, such that $A_{i_{j+1},i_j}\geq 0$ for all $0 \leq j<n-1$. The sequence ends because all the integers must be different from each other as mentioned in step-3, and the total number of basis states $D$ is finite.

Furthermore, we can show for any two different states $|\alpha_{i_j}\rangle, |\alpha_{i_k}\rangle \in \mathsf{\Gamma}_{0,i} = \{ |\alpha_{i_0}\rangle, |\alpha_{i_1}\rangle, ..., |\alpha_{i_n}\rangle \}$, the condition Eq.~(\ref{sign_free1}), \emph{i.e.} $A_{i_k i_j} \geq 0$, is satisfied. The reason is as follows: if $A_{i_k i_j} < 0$, then $A_{i_k i_j} \ \prod_{l=j}^{k-1} A_{i_{l+1},i_{l}} <0$ (assuming $j<k$ without loss of generality) which contradicts Eq.~(\ref{sign_free2}). As a result, all the states in $\mathsf{\Gamma}_{0,i}$ can be viewed as a single point in the state complex. This procedure does not lead to any contradiction: if for a state $|\alpha_l\rangle \in \mathsf{\Gamma}, |\alpha_l\rangle \notin \mathsf{\Gamma}_{0,i}$, there exist two different states $|\alpha_j\rangle, |\alpha_k\rangle \in \mathsf{\Gamma}_{0,i}$, such that $A_{il} \neq 0$ ($i=j,k$), then $A_{il}$ ($i=j,k$) must have the same sign because of Eq.~(\ref{sign_free2}). Note that in every loop of construction, when we need to add minus sign to a state $|\alpha_i\rangle$ in step-2, we should add minus signs to all the states belonging to the same point as $|\alpha_i\rangle$ in the state complex.

After one loop of construction, the number of points in the state complex is reduced by at least one. Because of the finiteness of the total number of basis states $D$, we will end up with a basis $\mathsf{\Gamma}_0$, which is obtained by adding minus signs to the basis states in $\mathsf{\Gamma}$, such that Eq.~(\ref{sign_free1}) is satisfied.


\section{Transformation of the Hamiltonian}
\label{Appen_transf_of_H}
In this appendix, we will discuss the transformation of the Hamiltonian under MST and PST in the fractionization formalism, which is more convenient to unveil the sign structure.

In the fractionization formalism, the creation operator of the spin-$S$ ``electron'' is fractionalized to a spin part and a charge part. We can use $2S+1$ kinds of spinons $d_{i,m}(m=-S,...,S)$ or two kinds of Schwinger bosons $b_{i\sigma}(\sigma=\uparrow(+1),\downarrow(-1))$ to represent the spin part of the particle at site $i$. The holon annihilation operator at site $i$ is denoted by $h_i$. The resulting two kinds of fractionalization formalisms are given by
\begin{eqnarray}\label{slave_particle1}
  c_{im}^\dagger &=& d_{i,m}^\dagger h_i,\\\label{slave_particle2}
  c_{im}^\dagger &=& \frac{1}{\sqrt{(S+m)!(S-m)!}}
    (b_{i\uparrow}^\dagger)^{S+m} (b_{i\downarrow}^\dagger)^{S-m} h_i.
\end{eqnarray}
We choose spinons $d_{i,m},b_{i\sigma}$ and holon $h_i$ all to be bosons. The local constraint on the Hilbert space at site $i$ is
\begin{equation}\label{constraint}
  n_i^h+\sum_m n_{i,m}^d=n_i^h+\frac{1}{2S}\left(n_{i\uparrow}^b+n_{i\downarrow}^b\right)=1,
\end{equation}
where $n_{i,m}^d=d_{i,m}^\dagger d_{i,m}, n_{i\sigma}^b=b_{i\sigma}^\dagger b_{i\sigma}, n_i^h=h_i^\dagger h_i$ are particle number operators of each boson.

Let us focus on the transformation of the $J$-term first. It is more convenient to use spinons $b_{i\sigma}$ to express the original $J$-term in Eq.~(\ref{t-J}):
\begin{equation}\label{HJ1_frac}
  H_J = \sum_{\langle ij\rangle,\sigma,\sigma'} -\frac{J_{ij}}{2} \left(\sigma\sigma'\right) b_{i\sigma}^\dagger b_{j\bar\sigma}^\dagger b_{j\bar\sigma'} b_{i\sigma'}.
\end{equation}
The Marshall sign transformation is a local unitary transformation which can be expressed as
\begin{eqnarray}\label{mst}
  U_{\mathrm{mst}} &=& \exp\left( i\pi\sum_{j,m} j (S+m) n_{j,m}^b \right).
\end{eqnarray}
We can also substitute $b_{j\sigma} \rightarrow (-\sigma)^j b_{j\sigma}$ to absorb the Marshall sign \cite{Marshall1955,Auerbach}, resulting in a $J$-term with trivial sign structure:
\begin{equation}\label{HJ1_after_Marshall}
  H_J' = U_{\mathrm{mst}} H_J U_{\mathrm{mst}}^\dagger = \sum_{\langle ij\rangle,\sigma,\sigma'} -\frac{J_{ij}}{2}\, b_{i\sigma}^\dagger b_{j\bar\sigma}^\dagger b_{j\bar\sigma'} b_{i\sigma'},
\end{equation}
which possesses only nonpositive signs in front of every term. This result indicates that MST already makes the $J$-term sign trivial. Under the nonlocal phase string transformation
\begin{eqnarray}\label{pst}
  U_{\mathrm{pst}} &=& \exp\left(i\pi\sum_{i<j;m}(S+m)n_i^h n_{j,m}^b\right),
\end{eqnarray}
the $J$-term is invariant:
\begin{equation}\label{HJ_afterU}
  \tilde H_J = U_{\mathrm{pst}} H_J' U_{\mathrm{pst}}^\dagger = H_J' = \sum_{\langle ij\rangle,\sigma,\sigma'} -\frac{J_{ij}}{2}\, b_{i\sigma}^\dagger b_{j\bar\sigma}^\dagger b_{j\bar\sigma'} b_{i\sigma'}.
\end{equation}
The above PST is the high-spin generalization of the spin-$1/2$ phase string transformation \cite{phase_1D_1,phase_1D_2,Weng1997,Wu2008sign}. The physical meaning of PST is to add a sign $(-1)^{S+m}$ to the basis states whenever a holon $h_i$ appears on the left side of a spinon $d_{j,m}$. The invariance of the $J$-term under PST comes from the fact that $H_J$ does not change the positions of holons, and $H_{i,i+1}^J$, when acting on the Hilbert space of two spin-1 particles on sites $i$ and $i+1$, does not change the number $n_{i,0}^d+n_{i+1,0}^d (\mathrm{mod}~2)$.

Now turn to the transformation of the $t$-term. It is more convenient to use spinons $d_{i,m}$ to express the $t$-term in Eq.~(\ref{t-J}):
\begin{equation}\label{}
  H^t = \sum_{i,m} -t_{ij} \left( d_{i,m}^\dagger d_{i+1,m} h_{i+1}^\dagger h_i +\mathrm{h.c.}\right).
\end{equation}
After the absorption of the Marshall sign by MST: $d_{j,m} \rightarrow (-1)^{(S+m)j} d_{j,m}$, the $t$-term becomes
\begin{equation}\label{Ht_after_Marshall}
  H_t' = \sum_{i,m} -t_{ij} \left((-1)^{S+m}
    d_{i,m}^\dagger d_{i+1,m} h_{i+1}^\dagger h_i +\mathrm{h.c.}\right).
\end{equation}
To keep track of the signs appearing when a hole exchanges with a spinon, we perform the nonlocal phase string transformation Eq.~(\ref{pst}) to the Marshall sign transformed $t$-term Eq.~(\ref{Ht_after_Marshall}). We then totally remove the signs of the $t$-term without adding new signs to the $J$-term:
\begin{equation}\label{Ht_afterU}
  \tilde H_t = U_{\mathrm{pst}} H_t' U_{\mathrm{pst}}^\dagger = \sum_{i,m} -t_{ij}
  \left(d_{i,m}^\dagger d_{i+1,m} h_{i+1}^\dagger h_i +\mathrm{h.c.}\right).
\end{equation}

The trivial sign structure of the $t$-$J$ Hamiltonian after MST and PST Eqs.~(\ref{HJ_afterU}) and (\ref{Ht_afterU}) is accompanied with physical operators with highly nontrivial nonlocal signs. As an example, the creation operator of the original spin-$S$ ``electron'' Eq.~(\ref{slave_particle1}) is now given by
\begin{widetext}
\begin{equation}\label{slave_particle3}
  \tilde c_{j,m}^\dagger = U c_{j,m}^\dagger U^\dagger
       =(-1)^{(S+m)j} \exp\left(i\pi\sum_{l<j;m} (S+m) n_l^h n_{j,m}^b\right) b_{j,m}^\dagger h_j \exp\left( i\pi\sum_{l>j;m} (S+m) n_j^h n_{l,m}^b \right),
\end{equation}
\end{widetext}
where $U=U_{\mathrm{pst}}\ U_{\mathrm{mst}}$ is the unitary transformation as a combination of MST and PST. For the spin-$1$ case, the above formulas can be written separately as
\begin{widetext}
\begin{eqnarray}\label{slave_particle4}
  \tilde c_{j,\pm1}^\dagger &=& b_{j,\pm1}^\dagger h_j
    \exp\left( i\pi\sum_{l>j} n_j^h n_{l,0}^b \right),\\
  \tilde c_{j,0}^\dagger &=&(-1)^{j} \exp\left(i\pi\sum_{l<j} n_l^h n_{j,0}^b\right)
    b_{j,0}^\dagger h_j \exp\left( i\pi\sum_{l>j} n_j^h n_{l,0}^b \right).
\end{eqnarray}
\end{widetext}
Any correlation function of the original ``electron'' operators contains the nontrivial nonlocal signs which may dramatically change the properties of the correlation function, although the ground state of Eqs.~(\ref{HJ_afterU}) and (\ref{Ht_afterU}) is simple in the sense of sign structure.

\section{Proofs of the generalized Marshall theorem and Lieb-Mattis theorem}
\label{Appen_theorems}
We will prove the two theorems stated in Sec.~\ref{subsec_GSD}. The crucial point is the lemma on the sign structure of the 1D spin-$S$ $t$-$J$ model discussed in Sec.~\ref{subsec_tJ_GS}. Similar results for the spin-$1/2$ model were proposed \cite{Xiang1992}. Since our paper mainly focuses on the spin-$1$ model, our results are more general in the sense that the two theorems are valid for arbitrary spin $S$.

\emph{Proof of Theorem-1}.
Let us replace $t_{ij}$ in Eq.~(\ref{t-J}) by ``$t \times t_{ij}$'' where the dimensionless parameter $t\in[0,1]$ (when $t=0$, the $t$-term vanishes; when $t=1$, it is recovered). We first investigate the $t$-$J$ model with $t=0$. The system now becomes a collection of antiferromagnetic Heisenberg chain segments separated by static holes. There are totally $C_L^{N_h}=L!/(N_h!N!)$ different hole distribution configurations denoted by $\{h_i\}$. The total Hilbert space $\mathcal{H}$ is a direct sum of $C_L^{N_h}$ subspaces which are disconnected under the action of the $t$= 0 Hamiltonian: $\mathcal{H} = \oplus_{\{h_i\}} \mathcal{H}(\{h_i\})$. As a result, the true ground state $|\mathsf{\Psi}_0\rangle$ in $\mathcal{H}$ is the one that has the lowest energy among the ground states in the subspaces $\mathcal{H}(\{h_i\})$.

For a fixed subspace $\mathcal{H}(\{h_i\})$, the ground state $|\mathsf{\Psi}_0(\{h_i\})\rangle$ can be expressed as the tensor product of $|\psi_j\rangle$, which is the ground state of the spin segment labeled by $j$ and satisfies the Marshall theorem. We will show that $|\mathsf{\Psi}_0\rangle$ is a state with all the spins forming a complete spin chain of length $N$ without breaking. To verify this result, let us begin with a general subspace and focus on two nearest-neighbor segments with lengths $N_1$ and $N_2$. Both the ground states $|\psi_1\rangle$ and $|\psi_2\rangle$ of the two segments have strictly positive coefficients in the Marshall basis. If we put the two segments together (labeling the ending site of the first segment by $I$, and the beginning site of another segment by $I+1$), the only relevant Hamiltonian term that potentially contributes to the energy change is $\Delta H=J_{I,I+1}\bold S_{I}\cdot\bold S_{I+1}$ defined on link $\langle I,I+1\rangle$. For the tensor product state $|\psi\rangle = |\psi_1\rangle\otimes|\psi_2\rangle$, the energy difference $\langle\psi | \Delta H | \psi\rangle$ is strictly negative because of the strictly positive nature of the coefficients of $|\psi_1\rangle$ and $|\psi_2\rangle$ in the Marshall basis. Therefore, by ignoring the holes at the two boundaries, the ground state $|\mathsf{\Psi}_0\rangle$ is the same as the ground state of a complete spin chain with length $N$ and no breaking. The degeneracy for $|\mathsf{\Psi}_0\rangle$ has two sources: the positions of the \emph{static} holes [$(N_h+1)$-fold degeneracy], and the possible spin degeneracy due to the Marshall theorem. In 1D, $|N_A-N_B|=0$ (or $1$)  if $L$ is even (or odd), thus these ground states have the total spin quantum number as claimed in the Theorem 1.

Now, we tune $t$ from zero to nonzero. As a discrete label of the irreducible representations of the spin rotational symmetry, the total spin quantum number of the ground states can not change, but the degeneracy due to the positions of the static holes may be lifted. Note that the states with spin degeneracy have different $S_{\mathrm{tot}}^z$, while the $N_h+1$ states degenerated due to hole placement have the same $S_{\mathrm{tot}}^z$. The lemma in Sec.~\ref{subsec_tJ_GS} states that the ground state is unique in the subspace with fixed $S_{\mathrm{tot}}^z$ for all $t>0$. Therefore, the ground-state degeneracy due to the hole placement is truly lifted. The ground state is unique apart from the $(2S_{\mathrm{tot}}^0+1)$-fold spin degeneracy.

\emph{Proof of Theorem-2}.
We also consider the case $t=0$ first. In fact, the arguments in the proof of Theorem-1 at $t=0$ are valid in the subspace $\mathcal{H} ( S^z_{\mathrm{tot}}=M )$ instead of the total Hilbert space $\mathcal{H}$. Thus we have the conclusion that the ground state of the $t$-$J$ model with $t=0$ in the subspace $\mathcal{H} ( S^z_{\mathrm{tot}}=M )$ is the same as that of a complete spin chain of length $N$ in the subspace $\mathcal{H} ( S^z_{\mathrm{tot}}=M )$. According to the Lieb-Mattis theorem \cite{Lieb-Mattis1962} on the energy ordering of spin systems, Theorem-2 is correct at $t=0$.

Let us now consider nonzero $t$. If Theorem-2 is broken for some nonzero $t$, say $E(S_{\mathrm{tot}}) \geq E(S_{\mathrm{tot}}+1)$ for some $S_{\mathrm{tot}} \geq S_{\mathrm{tot}}^0$, then by decreasing the parameter $t$, there exists a critical nonzero $t_c$, such that $E(S_{\mathrm{tot}}) = E(S_{\mathrm{tot}}+1)$ at this $t=t_c$ point. The ground state in the subspace $\mathcal{H} ( S^z_{\mathrm{tot}} = S_{\mathrm{tot}} )$ is now at least two fold degenerate. This conclusion contradicts the statement that the ground state is unique in the subspace $\mathcal{H} ( S^z_{\mathrm{tot}} = S_{\mathrm{tot}} )$ according to the Perron-Frobenius theorem. Therefore, Theorem 2 is valid for all positive $t$.

\section{Phase diagram}
\label{Appen_phase_diag}
\subsection{Charge sector}
Although we have asserted in Sec.~\ref{subsec_tJ_GS} that there is an energy gap for finite system due to the Perron-Frobenius theorem, it does not exclude the possibility that the energy gap shrink to zero in the thermodynamic limit. Take the antiferromagnetic spin-$S$ chain for example: the Marshall theorem tells us there is an energy gap for finite spin chain no matter whether $S$ is an integer or a half-odd-integer; but Haldane showed that the energy gap remains nonzero in the thermodynamic limit only for integer spin chains \cite{Haldane1983,Haldane1985}.

For the 1D $t$-$J$ model, when a hole breaks the spin chain into two segments, the exchange energy will be lifted by $2J_{\mathrm{eff}}$, where $-J_{\mathrm{eff}}$ (of order $J$) is the energy per bond of a Heisenberg spin chain. If two holes occupy nearby sites, the energy cost is only $J_{\mathrm{eff}}$. Effectively, holes will acquire a nearest-neighbor attractive interaction $-J_{\mathrm{eff}}$. Therefore, we can write the effective charge model
\begin{equation}\label{charge}
  H_{\mathrm{eff}} = -t \sum_i \left( h_i^\dagger h_{i+1} + \mathrm{H.c.} \right) - J_{\mathrm{eff}} \sum_i n_i^h n_{i+1}^h,
\end{equation}
which is a spinless fermion model with nearest neighbor attractive interaction. Note that the above effective charge model is valid no matter whether the initial spin chain is gapless or gapped.

The effective charge model [Eq.~(\ref{charge})] can be well understood since the spin-$1/2$ ferromagnetic $XXZ$ model
\begin{equation}\label{XXZ}
  H_{XXZ} = \sum_{\langle ij \rangle} \left[ -J_{XY} (S_i^x S_j^x +S_i^y S_j^y) - J_z S_i^z S_j^z \right]
\end{equation}
can be mapped exactly to Eq.~(\ref{charge}) by using Jordan-Wigner transformation \cite{JW1928,Giamarchi2004}. The parameters of the effective charge model and the ferromagnetic $XXZ$ model are related by
\begin{equation}
  J_{XY}=2\,t, \quad J_z=J_\mathrm{eff}.
\end{equation}
Note that the chemical potential term, which can be used to tune the number of holes in Eq.~(\ref{charge}), corresponds to the magnetic filed term in the $XXZ$ model Eq.~(\ref{XXZ}).

The phase diagram of the $XXZ$ model is obtained from the Bethe ansatz or bosonization method \cite{Haldane1980,Giamarchi2004}. The ferromagnetic phase of the $XXZ$ model in the region $J_z>J_{XY}$ implies the phase separation of the effective charge model when $t/J_{\mathrm{eff}} < 1/2$. On the other hand, for $J_z<J_{XY}$ or $t/J_{\mathrm{eff}} > 1/2$, the effective charge model will behave as a Luttinger liquid. These results suggest the charge sector of the $t$-$J$ model possesses a phase transition from gapped phase to gapless Luttinger liquid phase.

If we choose open boundary conditions, holes have the tendency to live at the boundary to avoid energy cost of order $J_{\mathrm{eff}}$ by breaking the spin chain. On the other hand, the energy gain by injecting a hole into the mobile band is $2\,t$. Therefore, we expect that holes will be localized at the boundary (gapped) if $2\,t<J_{\mathrm{eff}}$ and delocalized (gapless) if $2\,t>J_{\mathrm{eff}}$, which agrees with the discussion above. In Appendix~\ref{Appen_single_hole}, we illustrate this picture in the single-hole case.

Note that, for the $XXZ$ model, the phase transition point $J_z=J_{XY}$ does not change with respect to the magnetization. However, the critical value of $t/J$ depends on the hole doping concentration $\delta$, since parameter $J_{\mathrm{eff}}$ in the effective charge model [Eq.~(\ref{charge})] will vary with respect to $\delta$ generally.

\subsection{Spin sector}
The spin sector of our model is clear in two limit cases: $t/J \rightarrow 0$ and $t/J \rightarrow \infty$. These two cases will be discussed separately.

As claimed above, the $t$-$J$ model possesses phase separation for small $t/J$. As we decrease $t/J$, the system will adiabatically connect to the pure spin chain, as holes will be localized at two boundaries. Therefore, the small $t/J$ phase of the spin-$S$ $t$-$J$ model is in the same phase as the spin-$S$ antiferromagnetic Heisenberg model. The spin sector is gapless if $S$ is half-odd-integer, and gapped if $S$ is integer \cite{Haldane1983,Haldane1985}.

It is known that, for the spin-$1/2$ $t$-$J$ model, the spin sector and the charge sector are totally separated in the limit $t/J \rightarrow \infty$ \cite{large_t_phase}. The charge degrees of freedom behave as spinless fermions, while the spin sector is equivalent to the 1D spin-$1/2$ Heisenberg model. Therefore, the spin sector of the spin-$S$ $t$-$J$ model in the limit $t/J \rightarrow \infty$ is the same as in the limit $t/J \rightarrow 0$. This is also true for the $t$-$J$ model with other spin $S$. The reason is as follows:

Let the ground state of the spin-$S$ $t$-$J$ model with $J=0$ to be
\begin{equation}\label{}
  |\psi_c\rangle = \sum_{\{i_n\}} a(\{i_n\}) \prod_{n=1}^N c_{i_n,m_n}^\dagger |0\rangle,
\end{equation}
where $n$ labels the spin-$S$ ``electrons'', and $i_n, m_n$ are the position and spin of the $n$-th ``electron'' correspondingly. The ground states are highly degenerated as the spin configuration $\{m_n\}$ can be chosen arbitrarily. The ground state of the $t$-$J$ model with an infinitesimally small $J$ must have the same $t$-term energy as $|\psi_c\rangle$, otherwise the energy gain from the $J$-term (note that $J \rightarrow 0$) can not afford the energy cost of the $t$-term. Therefore, the only possible ground state is the one with spin-charge separation:
\begin{equation}\label{}
  |\psi\rangle = \sum_{\{i_n\}} a(\{i_n\}) \left( \sum_{\{m_n\}} b(\{m_n\}) \prod_{n=1}^N c_{i_n,m_n}^\dagger |0\rangle \right).
\end{equation}
To further gain $J$-term energy, the coefficients $b(\{m_n\})$ must be the same as the ground state of the pure Heisenberg spin chain:
\begin{equation}\label{}
  |\psi_s\rangle =  \sum_{\{m_n\}} b(\{m_n\}) \prod_{n=1}^N c_{n,m_n}^\dagger |0\rangle.
\end{equation}
The spin-charge separation truly happens in the limit $t/J \rightarrow \infty$ for the $t$-$J$ model with arbitrary spin-$S$.

We can construct an exactly solvable fixed point Hamiltonian to illustrate the spin-charge separation in the $t/J \rightarrow \infty$ limit. The fixed-point wave function of the pure spin model, which shares the same universal properties with the ground state of the spin-$1$ Heisenberg chain, is the so-called AKLT state \cite{AKLT1987,AKLT1988}. By replacing the $J$-term of the 1D spin-$1$ $t$-$J$ model by the AKLT Hamiltonian, we obtain the doped AKLT model. The ground state of this model is explicitly constructed in Ref.~\onlinecite{Zhang1989}. We will discuss more about this wave function and calculate the string order parameter for the ground state of the doped AKLT model in Sec.~\ref{subsec_string_order}.

\subsection{Single-hole $t$-$J$ model}
\label{Appen_single_hole}

\begin{figure}[b]
\includegraphics[width=8.5cm]{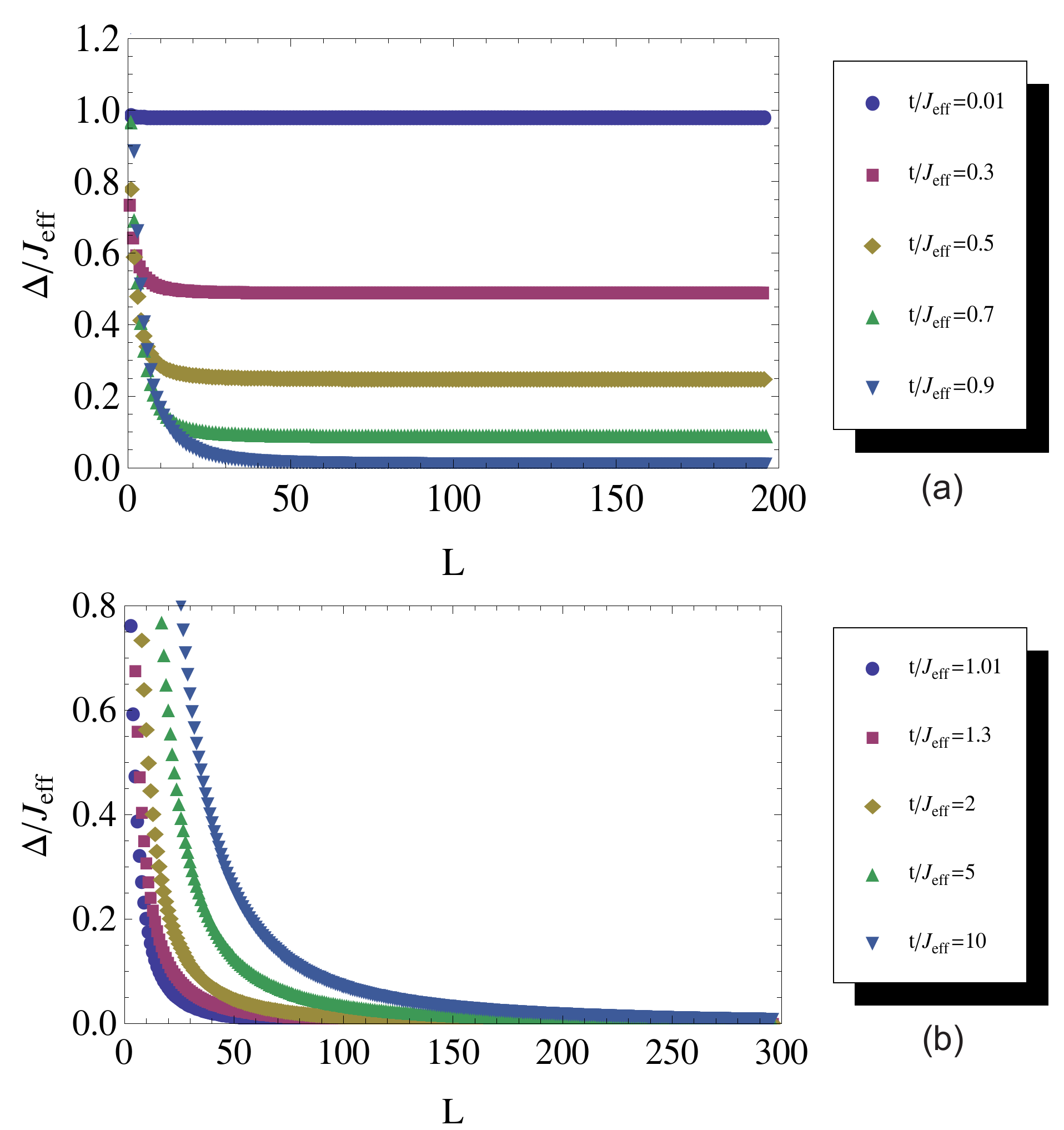}
\caption{\label{fig:gap_L} (Color online) The energy gap of effective charge model as a function of lattice size $L$. (a) For $t/J_{\mathrm{eff}}<1$, the system has a finite gap which decreases with the increasing of $t/J_{\mathrm{eff}}<1$. (b) For $t/J_{\mathrm{eff}} \geq 1$, the system is always gapless in the thermodynamic limit.}
\end{figure}

\begin{figure}[b]
\includegraphics[width=8.5cm]{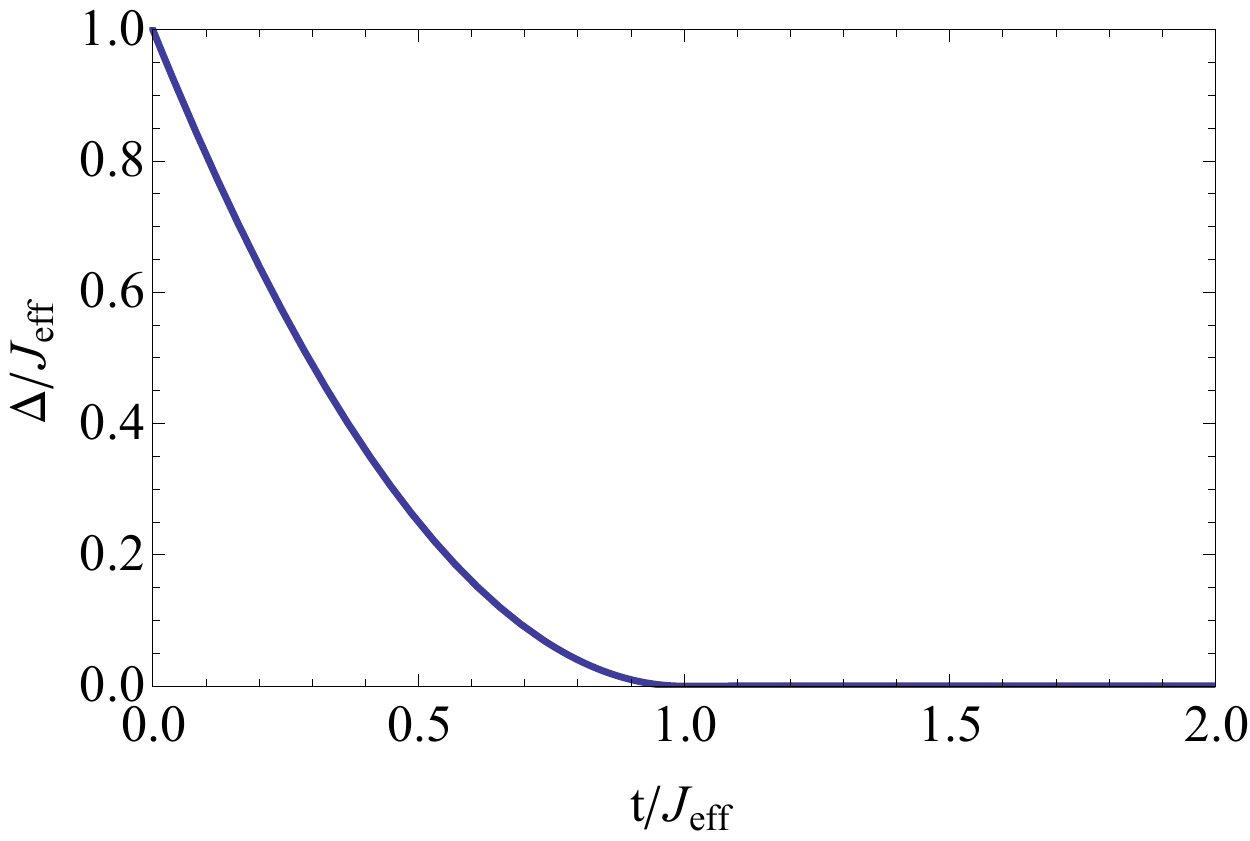}
\caption{\label{fig:gap_t} The charge gap $\Delta/J_{\mathrm{eff}}$ as a function of $t/J_{\mathrm{eff}}$. There is a quantum critical point at $t/J_{\mathrm{eff}}=1$ which separates the charge gapped phase with the charge gapless phase. The energy gap is measured in a system with $L=500$ lattice sites.}
\end{figure}

In this appendix, we will derive an effective model for the charge sector of the \emph{single-hole} $t$-$J$ model. Numerical and analytical calculation both show that there is a critical value for $t/J$: if $t/J$ is small, the hole will be localized at the boundary of the chain and there is a finite-energy gap in thermodynamic limit; if $t/J$ is large, the hole will be extended in space and the system is gapless.

Let $E_0(L)$ be the ground-state energy for an antiferromagnetic Heisenberg spin-$1$ chain with $L$ lattice sites. In the limit $L \rightarrow \infty$, the total energy approaches $E_0(L) \rightarrow -J_{\mathrm{eff}}(L-1)$, where $-J_{\mathrm{eff}}$ is the exchange energy per bond, which is of course proportional to the original $J$ in the Heisenberg model. Now, consider a system with $L$ lattice sites and a static hole ($t=0$) at site $n$ ($1<n<L$). The true ground state $|\psi_0(n)\rangle$ is the tensor product of the ground states of two spin chains with length $L_1=n-1$ and $L_2=L-n$. The ground-state energy is $E_0(L_1,L_2) = E_0(L_1) + E_0(L_2)$, which becomes $-J_{\mathrm{eff}}(L_1+L_2-2) = -J_{\mathrm{eff}}(L-3) = E_0(L) + 2J_{\mathrm{eff}}$ in the limit $L_1,L_2 \rightarrow \infty$.
Consider now another wave function $|\psi_1(n)\rangle$, which is obtained from the ground state of a spin chain with length $L-1$, by adding a lattice site at site $n$ and shifting sites greater than $n$ by one. The total exchange energy for $|\psi_1(n)\rangle$ is $E_1(L_1,L_2) = [(L-3)/(L-2)]E_0(L-1)$. In the limit $L_1+L_2 \rightarrow \infty$, $E_1(L_1,L_2)$ approaches also $-J_{\mathrm{eff}}(L-3)$. Thus, we can conclude that the wave function $|\psi_1(n)\rangle$ is a good approximation for the ground state of a spin chain with a static hole at site $n$ in the limit $L_1,L_2 \rightarrow \infty$. The energy difference between this state and the ground state of a spin chain with length $L$ is $2J_{\mathrm{eff}}$. We should also mention that if $n=0$ or $L$, the energies for $|\psi_0\rangle$ and $|\psi_1\rangle$ are both $-J_{\mathrm{eff}}(L-2) = E_0(L) + J_{\mathrm{eff}}$, which is lower than that of the $1<n<L$ cases since the hole on the boundary does not break the spin chain.

The advantages for using $|n\rangle \equiv |\psi_1(n)\rangle$ rather than $|\psi_0(n)\rangle$ is that the subspace $\{|n\rangle\}$ is closed under the action of the hopping term $H_t$:
\begin{equation}\label{}
  H_t | n \rangle = -t \left( |n-1\rangle + |n+1\rangle \right).
\end{equation}
This implies spin-charge separation as the hole hopping process does not modify the spin background. The above spin-charge separation is exact rather than an approximation if we are dealing with the doped AKLT model rather than the $t$-$J$ model \cite{Zhang1989}.

After combining the exchange energy and the hopping term together, we get the effective charge model for a single hole
\begin{equation}\label{H_eff1}
  H_{\mathrm{eff}} = -t\sum_{n=1}^{L-1} \left( |n\rangle \langle n+1| + \mathrm{H.c.} \right) - J_{\mathrm{eff}} \left( |1\rangle\langle1| + |L\rangle\langle L| \right),
\end{equation}
up to some constant. We have made an approximation that the exchange energies for different states $|n\rangle$ ($1<n<L$) are the same: $E_0(L) + 2J_{\mathrm{eff}}$. This model can be solved easily numerically since the Hilbert space dimension is $L$, rather than an exponential function of $L$ as the original $t$-$J$ model. The energy gap $\Delta$, as a function of lattice length $L$ and $t/J_{\mathrm{eff}}$, is shown in Figs.~\ref{fig:gap_L} and \ref{fig:gap_t}, respectively. These results tell us the energy gap is nonzero in thermodynamic limit if $t<J_{\mathrm{eff}}$, and zero if $t \geq J_{\mathrm{eff}}$.

In fact, the above conclusion can be drawn analytically. First let us consider the point $t=J_{\mathrm{eff}}$. We will set $t=J_{\mathrm{eff}}=1$ for simplicity. The Hamiltonian at this point has the following properties: all off-diagonal elements are $-1$ or $0$; the sum of every column of the Hamiltonian is zero (after adding a constant 2 to the Hamiltonian). This is an example of the \emph{Laplacian models} defined in Sec.~\ref{subsec_geo}. 
The ground state of the Laplacian model can be solved exactly: an equal weight superposition of all basis states in an irreducible space. In our specific case, the ground state is
\begin{equation}\label{}
  | \phi ( t=J_{\mathrm{eff}} ) \rangle = \sum_n |n\rangle,
\end{equation}
which indicates that the hole density distribution will be uniform along the whole chain. The system is gapless since the ground-state energy $-2$ touches the bottom of the energy band for the hopping term $\epsilon_k = -2 \cos k$.

If $t<J_{\mathrm{eff}}$, the onsite potential at site $1$ and $L$ is lower than that of the Laplacian model. Since the hole has uniform density in the ground state of the Laplacian model, we expect the hole density distribution now will be centered at the two boundaries. In fact, the ground state in which the hole is localized at the left boundary of the chain, in the thermodynamic limit, is
\begin{equation}\label{localized_state}
  | \phi ( t<J_{\mathrm{eff}} ) \rangle = \sum_n \left( \frac{t}{J_{\mathrm{eff}}} \right)^n |n\rangle,
\end{equation}
with proper normalization factor. A similar result holds for another ground state in which the hole is localized at the right boundary. The ground-state energy with respect to Eq.~(\ref{H_eff1}) is $E=-J_{\mathrm{eff}}-t^2/J_{\mathrm{eff}}$, which is always below the bulk energy band. Therefore, there is a charge energy gap in the thermodynamic limit for $t<J_{\mathrm{eff}}$, as indicated by Fig.~\ref{fig:gap_L}(a).

On the other hand, for $t>J_{\mathrm{eff}}$, the energy gap is zero in thermodynamic limit [see Fig.~\ref{fig:gap_L}(b)]. The onsite boundary potential is too high, and every eigenstate is extended in space . It can also be seen from the fact that the localized state [Eq.~(\ref{localized_state})] can not be normalized if $t/J_{\mathrm{eff}}>1$.


\section{String order parameter for doped AKLT}
\label{Appen_stringorder}
Let us first calculate the string order parameter for the AKLT state. The calculation for the doped AKLT state is parallel to this procedure. Up to a global normalization factor, the AKLT state Eq.~(\ref{AKLT_wf}) can be written as
\begin{equation}\label{AKLT_wf2}
  |\Psi_{\mathrm{AKLT}}\rangle = \sum_{\{m_i\},\; m_i=0,\pm1} \delta^{\mathrm{HAF}}_{\{m_i\}}\ \eta_M \left( \frac{1}{\sqrt 2} \right) ^{N_0} | \{ m_i \} \rangle.
\end{equation}
The coefficient of a given Ising basis state $| \{ m_i \} \rangle$ has three contributions: (i) $\delta^{\mathrm{HAF}}_{\{m_i\}}$ is a factor to ensure the \emph{hidden antiferromagnetic order} of the Ising configuration $\{ m_i \}$: if the $m_i=\pm1$ spins form an antiferromagnetic chain after ignoring the $m_i=0$ spins, then $\delta^{\mathrm{HAF}}_{\{m_i\}} = 1$; otherwise, $\delta^{\mathrm{HAF}}_{\{m_i\}} = 0$. (ii) The sign factor $\eta_M = \left(-1\right)^{N_0^B}$ is the Marshall sign where $N_0^B$ is the number of $m_i=0$ spins at sites belonging to the $B$ sublattice. (iii) The factor $\left(1/\sqrt 2\right)^{N_0}$, where $N_0$ is the number of $m_i=0$ spins, comes from the normalization of Schwinger boson representation of spin-$1$: $|m_i=0\rangle = b_{i\uparrow}^\dagger b_{i\downarrow}^\dagger |0\rangle_i,\quad |m_i=\pm1\rangle = \left(1/\sqrt 2\right) \left( b_{i,\uparrow/\downarrow} ^\dagger \right)^2 |0\rangle_i$.
We can choose one of the four ground states to calculate the string order parameter, by assuming the first and the last $m_i \neq 0$ on the chain are both $m_i=1$.
To calculate the string order parameter, we should perform the so-called string order transformation $U_{\rm sot}$ \cite{deNijs1989,Girvin1989}
\begin{equation}\label{string_order_transf}:
  U_{\mathrm{sot}} = \prod_{i<j} \exp \left( i\pi S_i^z S_j^x \right).
\end{equation}
It is also a nonlocal unitary transformation as PST. The string order transformation has two effects: (i) removing the Marshall signs; (ii) transforming the hidden antiferromagnetic order to the hidden ferromagnetic order. After this transformation, the AKLT state Eq.~(\ref{AKLT_wf2}) becomes
\begin{equation}\label{AKLT2}
  | \tilde \Psi_{\mathrm{AKLT}}\rangle = \sum_{\{m_i\},\; m_i=0,\pm1} \delta^{\mathrm{HFM}}_{\{m_i\}} \ \left( \frac{1}{\sqrt 2} \right) ^{N_0} | \{ m_i \} \rangle,
\end{equation}
where $\delta^{\mathrm{HFM}}_{\{m_i\}} = 1$, if the $m_i \neq 0$ spins form a ferromagnetic chain with all $m_i=1$ after ignoring the $m_i=0$ spins; otherwise, $\delta^{\mathrm{HFM}}_{\{m_i\}} = 0$. The original string order parameter
\begin{equation}\label{SOP}
O_{\mathrm{string}} = \lim_{|i-j| \rightarrow \infty} \left\langle -S_i^z \exp\left( i \sum_{i<l<j} \pi S_l^z \right) S_j^z \right\rangle
\end{equation}
is transformed to the usual ferromagnetic correlation function
\begin{equation}\label{SOP2}
\tilde O_{\mathrm{string}} = U_{\mathrm{sot}} \ O_{\mathrm{string}} \ U_{\mathrm{sot}}^\dagger = \lim_{|i-j| \rightarrow \infty} \left\langle S_i^z S_j^z \right\rangle.
\end{equation}
\begin{widetext}

The calculation of the string order parameter Eq.~(\ref{SOP2}) for the AKLT state Eq.~(\ref{AKLT2}) is now straightforward:
\begin{eqnarray}\label{}\nonumber
\tilde O_{\mathrm{string}}^{\mathrm{AKLT}} &=& \lim_{|i-j| \rightarrow \infty} \frac{ \langle \tilde \Psi_{\mathrm{AKLT}} | S_i^z S_j^z | \tilde \Psi_{\mathrm{AKLT}} \rangle } { \langle \tilde \Psi_{\mathrm{AKLT}} | \tilde \Psi_{\mathrm{AKLT}} \rangle }\\\nonumber
&=& \lim_{|i-j| \rightarrow \infty} \frac { \sum_{\{m_i\},\; m_i=0,\pm1} \delta^{\mathrm{HFM}}_{\{m_i\}} \ (1/2)^{N_0} m_i m_j } { \sum_{\{m_i\},\; m_i=0,\pm1} \delta^{\mathrm{HFM}}_{\{m_i\}} \ (1/2)^{N_0} }\\
&=& \left( \frac{2}{3} \right)^2.
\end{eqnarray}
In the last step, the statistical averages for $m_i$ and $m_j$ are decoupled. Each of them contributes a factor $2/3$, because the factor $(1/2)^{N_0}$ makes the statistical distributions for $m_i=0$ and $m_i=1$ are $1/3$ and $2/3$.

Now turn to the string order parameter for the doped AKLT state. By replacing the $J$-term of the 1D spin-$1$ $t$-$J$ model by the AKLT Hamiltonian (\ref{AKLT_H}), we obtain the doped AKLT model (``\emph{d}AKLT'')
\begin{equation}
    H_{\mathrm{\emph{d}AKLT}} = -t\sum_{i,m}\left(c_{i,m}^\dagger c_{i+1,m}+\mathrm{h.c.}\right) + J\sum_{i} \left(\mathbf S_i\cdot \mathbf S_{i+1} +\frac{1}{3}\left(\mathbf S_i\cdot \mathbf S_{i+1}\right)^2 + \frac{2}{3} n_i n_{i+1} \right).
\end{equation}
The ground state of this model is explicitly constructed in Ref.~\onlinecite{Zhang1989}. Note that the constant term $2/3$ in Eq.~(\ref{AKLT_H}) is replaced by $(2/3) n_i n_{i+1}$, which is now one of the crucial points to make the model exactly solvable. The ground state of this model with $N_h$ holes is given by \cite{Zhang1989}
\begin{equation}\label{dAKLT}
  | \Psi_{\mathrm{\emph{d}AKLT}}(N_h)\rangle = \sum_{\{h_i\},P\in S_{N_h}} \mathrm{sgn}(P) \ \exp\left(i\sum_j k_{P_j} h_j\right) \sum_{\{m_i\},\; m_i=0,\pm1} \delta^{\mathrm{HAF}}_{\{m_i\}} \ \eta'_M \ \left( \frac{1}{\sqrt 2} \right) ^{N_0} | \{h_i\};\{ m_i \} \rangle,
\end{equation}
where $\mathrm{sgn}(P)$ is the signature of the element $P$ of the permutation group $S_{N_h}$, $\{k_j\}$ are $N_h$ momenta with lowest single particle energies \cite{Zhang1989}. Note that for the Marshll sign $\eta'_M$ in the above formula, the sublattice $B$ is defined on a lattice by ignoring all hole sites. After string order transformation Eq.~(\ref{string_order_transf}), this state becomes
\begin{equation}\label{dAKLT2}
  | \tilde \Psi_{\mathrm{\emph{d}AKLT}}(N_h)\rangle = \sum_{\{h_i\},P\in S_{N_h}} \mathrm{sgn}(P) \ \exp\left(i\sum_j k_{P_j} h_j\right) \sum_{\{m_i\},\; m_i=0,\pm1} \delta^{\mathrm{HFM}}_{\{m_i\}} \ \left( \frac{1}{\sqrt 2} \right) ^{N_0} | \{h_i\};\{ m_i \} \rangle.
\end{equation}
We can perform similar calculation for the string order parameter:
\begin{eqnarray}\label{}\nonumber
\tilde O_{\mathrm{string}}^{\mathrm{\emph{d}AKLT}} &=& \lim_{|i-j| \rightarrow \infty} \frac{ \langle \tilde \Psi_{\mathrm{\emph{d}AKLT}} (N_h) | S_i^z S_j^z | \tilde \Psi_{\mathrm{\emph{d}AKLT}} (N_h) \rangle } { \langle \tilde \Psi_{\mathrm{\emph{d}AKLT}} (N_h) | \tilde \Psi_{\mathrm{\emph{d}AKLT}} (N_h) \rangle }\\\nonumber
&=& \lim_{|i-j| \rightarrow \infty} \frac { \sum_{\{h_i\}} \sum_{\{m_i\}} \delta^{\mathrm{HFM}}_{\{m_i\}} \left| \sum_{P\in S_{N_h}} \mathrm{sgn}(P) \ \exp\left(i\sum_j k_{P_j} h_j\right) \right|^2 \ (1/2)^{N_0} m_i m_j } { \sum_{\{h_i\}} \sum_{\{m_i\}} \delta^{\mathrm{HFM}}_{\{m_i\}} \left| \sum_{P\in S_{N_h}} \mathrm{sgn}(P) \ \exp\left(i\sum_j k_{P_j} h_j\right) \right|^2 \ (1/2)^{N_0} }\\
&=& \left( \frac{2}{3} \right)^2 (1-\delta)^2.
\end{eqnarray}
\end{widetext}
In the last step, we used the density-density correlation function for Fermi gas $\langle n_i n_j \rangle = \bar n ^2$ if $|i-j| \rightarrow \infty$. The Fermi gas behavior of the charge degrees of freedom contributes a factor $(1-\delta)^2$ besides the factor $(2/3)^2$ inherited from the AKLT state.


\end{document}